\documentclass{iopart}
\usepackage{graphicx}
\usepackage{amsfonts}
\usepackage[english]{babel}
\usepackage{times}
\usepackage{graphicx}

\newcommand{\re}[1]{\mathrm{Re}\left[#1\right]}
\newcommand{\im}[1]{\mathrm{Im}\left[#1\right]}
\renewcommand{\imath}[0]{\mathrm{i}}
\newcommand{\abs}[1]{\left\vert#1\right\vert}
\newcommand{\nt}[1]{#1_{\mathbf{k}}^p}

\newcommand{\psum}[0]{{\kern 4.5ex\raisebox{0.25ex}{$'$}\kern 
-4.5ex}\sum}

\begin{document}

\title{Casimir energy and entropy between dissipative mirrors}

\author{F.~Intravaia and C.~Henkel}
\address{Universit\"{a}t Potsdam, Institut f\"{u}r Physik, Am Neuen 
Palais 10, 14469 Potsdam, Germany}

\ead{francesco.intravaia@qipc.org}

\begin{abstract}
We discuss the Casimir effect between two identical, parallel
slabs, emphasizing the role of dissipation and temperature.  
Starting from quite general assumptions, we analyze the behavior of the Casimir
entropy in the limit $T\rightarrow 0$ and link it to the behavior
of the slab's reflection coefficients at low frequencies.
We also derive a formula in terms of a sum over modes, 
valid for dissipative slabs that can be interpreted in terms of a damped
quantum oscillator.
% \\
% Date: 22 Oct 07.
 \end{abstract} 

\section{Introduction}

In recent years, the Casimir force \cite{Casimir48} has gained the status of a
mesoscopic force, attracting considerable attention from 
physicists and engineers who work with nano- or 
micro-electromechanical devices \cite{Mohideen98, Bressi02, Chan01}.
In particular in view of these applications, many theoretical efforts
are addressing more realistic configurations. But also for 
fundamental applications, a precise knowledge of experimentally 
relevant imperfections is valuable to tighten upper limits~%
\cite{Lambrecht03,Onofrio06}.
The Casimir effect originally describes the force 
between two parallel, perfectly reflecting slabs placed in vacuum 
at zero 
temperature, and it is quite clear that the generalization to a realistic 
experimental setup roughly articulates in three steps: 
(i) realistic materials, (ii) non-zero temperature, and (iii) other 
geometries. Step (i) was essentially achieved by Lifshitz in 
1955 \cite{Lifshitz56}, and within the same theory, addressing step
(ii) was apparently trivial. Surprisingly, up 
to now, the role of dissipation in the 
temperature correction to the force between two dissipative metallic 
slabs is unclear and at the center of hottest controversy around the 
Casimir effect~\cite{Bostrom00,Klimchitskaya01,Bezerra02,Svetovoy03,%
Brevik03,Mostepanenko06,Sernelius06a,Svetovoy06}. 
Lifshitz adopts a ``matter'' point of view to calculate the 
Casimir effect building on a stochastic (Langevin) version of the 
Maxwell equations where quantum fluctuating currents are responsible 
for the fields. Although the inclusion of dissipation is
quite natural in this theory, it becomes rather complicated
when we adopt a ``field'' point of view. The problem is that the
eigenfrequencies of the field, $\omega_{m}$, become 
complex with absorbing mirrors, so that the expression used by 
Casimir
\begin{equation}
\label{eq:CasEn}
E=\frac{\hbar}{2}
\left[\sum_{m}\omega_{m}\right]^{L}_{\infty}
\end{equation}
is meaningless. We use
the symbol $\left[\cdots\right]^{L}_{\infty}$ to denote the
renormalization of the sum over mode branches $m$, subtracting the
limiting value of the square bracket when the mirror distance $L \to
\infty$. Casimir's expression~(\ref{eq:CasEn}) 
is directly generalizable to partially reflecting (dispersive) 
mirrors~\cite{Schram73}, but it reveals to 
be rather tricky in the dissipative case. Among others,
Langbein~\cite{Langbein70, Langbein74} has suggested that one should take the real 
part of the complex energy, but his heuristic arguments remain 
unsatisfactory. 

In this paper we discuss two different aspects of 
including dissipation into the Casimir effect. 
We start with a rather general viewpoint on the Casimir entropy
at finite temperature, recover previous results and give a
criterion to classify them.
We also show how Eq.(\ref{eq:CasEn}) must be modified when
complex frequencies enter the theory, in order to take
consistently into account the dissipation phenomena.

\section{The Casimir entropy}

The current controversy arises when calculating 
the correction to the Casimir force at finite (nonzero) temperature. 
Although the calculation 
was implemented in Lifshitz's pioneering paper and the 
problem was already recognized by Schwinger et al.~\cite{Schwinger78},
it has been carefully examined only recently.
An orthodox application of Lifshitz's formula for metal has lead to 
some doubts concerning the way one describes their optical properties. 
Indeed, using the Drude model to describe the optical response of a 
metallic mirror leads to a completely different behavior of the 
TE and TM polorizations, in stark 
contrast to perfect reflectors (Dirichlet boundary conditions)
that, by definition, perfectly reflect 
both polarizations. This 
phenomenon is fairly striking at high temperatures, but
it leads to significant thermal corrections also, in particular
for the Casimir entropy, in the low temperature limit.
In fact, one important thermodynamical principle (the 
Nernst heat theorem) states that when we deal with an equilibrium 
situation the system entropy must go to zero in the limit $T \to 0$.

The final result of Lifshitz's paper is an expression for
the Casimir free energy as a sum over Matsubara frequencies:
\begin{equation}
\mathcal{F}=\frac{\hbar}{2\pi}\sum_{p,\mathbf{k}} 
\psum\limits_{n=0}^{\infty}\ \tau\ 
g^p_k(\imath n \tau,\tau)
\end{equation}
where $p$ labels the polarization (TM, TE) of an 
electromagnetic wave with wavevector $\mathbf{k}$ parallel to the 
slab surface. The primed sum has a prefactor $1/2$ in the first term, 
as usual. The function 
$g^p_k(\omega,\tau) = \ln\left[1-r^p_\mathbf{k}(\omega,\tau)^{2}
e^{-2\kappa L}\right]$ and we use [$\tau$ is the first Matsubara
frequency]
\begin{equation}
\label{eq:scaled-tau}
\kappa^2 = k^2 - \omega^2/c^2, \qquad 
\tau = 2 \pi k_B T /\hbar
\end{equation}
Note that we have allowed for a dependence on temperature of 
the reflection coefficients $r^p_\mathbf{k}(\omega,\tau)$. 
The Casimir entropy can be 
derived  from the previous expression by differentiation. 
Because the series converges uniformly in the 
parameter $\tau$ we can 
interchange
the sum and the derivative. In the 
following, we drop the sum over
the quantum numbers $p$ 
and $\mathbf{k}$ and set 
$k_{B} = 1$:
\begin{equation}
\label{eq:entroform}
S = - \frac{ 2 \pi }{ \hbar }
\frac{d\mathcal{F}}{d\tau} = - \psum\limits_{n=0}^{\infty}\ \left[
g(\imath n \tau,\tau) + \imath n\tau g_\omega(\imath n\tau, 
\tau) + \tau g_{\tau}(\imath n\tau,\tau) \right]
\end{equation} 
where we have defined
\begin{equation}
g_\omega(\omega,\tau) \equiv \partial_\omega
g(\omega,\tau)\quad\mbox{and}\quad 
g_{\tau}(\omega,\tau) \equiv
\partial_{\tau}g(\omega,\tau)
\end{equation}
Of course the first two terms in~(\ref{eq:entroform}) 
are more significant than the third, which 
is identically zero when the reflection coefficient does not depend 
on temperature. We ignore the third term for the moment and
suppress the second argument for simplicity.
The sum over $n$ is re-arranged as
\begin{equation}
\label{sum}
\sum_{n=0}^{\infty} 
\imath n\tau \left[g_\omega(\imath n \tau) - 
\frac{g(\imath n \tau)-g(\imath (n-1) \tau)}{\imath 
\tau}\right]
-\frac{1}{2}g(0) 
\nonumber
\end{equation}
Inside the brackets, we use the Taylor expansion of $g$ to get 
\begin{equation}
\label{expansion}
g(\imath (n-1) \tau)-g(\imath n\tau)+\imath\tau g_\omega(\imath 
n\tau)=
\sum_{k \ge 2} g_{\omega^k}(\imath n \tau)
\frac{(-\imath\tau)^k}{k!} 
\end{equation}
Therefore the entropy per mode can be rewritten as
\begin{equation}
S=-
\left(\sum_{n \ge 0, \, k \ge 2}n
g_{\omega^k}(\imath n \tau)\frac{(-\imath\tau)^k}{k!}- 
\frac{1}{2}g(0)\right)
\end{equation} 
Using the analytical properties of the function $g$, we can
express the $k$'th derivative in $\omega$ by a contour
integral along a path $C$ in the upper half complex plane
(the path must circle around each
$z = \imath n \tau$)
and get
\begin{equation}
\label{eq:entro}
S=- \left(\oint_{C}\sum_{n \ge 0, \,k \ge 2}\frac{g(z) n(-\imath\tau)^k}{(z-\imath n \tau)^{k+1}}  
\frac{dz}{2\pi\imath}  - \frac{1}{2}g(0)\right)
\end{equation} 
Exploiting the Euler-MacLaurin formula~\cite{Abramowitz71} 
we have [$B_{m}$ are the Bernoulli numbers]
\begin{equation}
\label{series}
\fl \sum_{n \ge 0}\frac{n(-\imath\tau)^k}{(z-\imath 
n\tau)^{k+1}} = \frac{(-\imath\tau)^{k-2}}{z^{k-1}}
\left[\frac{1}{k(k-1)}+\sum_{m \ge 1}B_{2m}\frac{(k+2m-1)!}{(2m-1)!k!}
\left(\frac{\tau}{z}\right)^{2m}\right]
\end{equation}
When this is substituted into Eq.(\ref{eq:entro}), we get 
an exact series expansion of the Casimir entropy in terms of 
the temperature and of the derivatives of $g(z)$ at $z=0$. 

\begin{figure}
\centerline{
\includegraphics*[width=8cm]{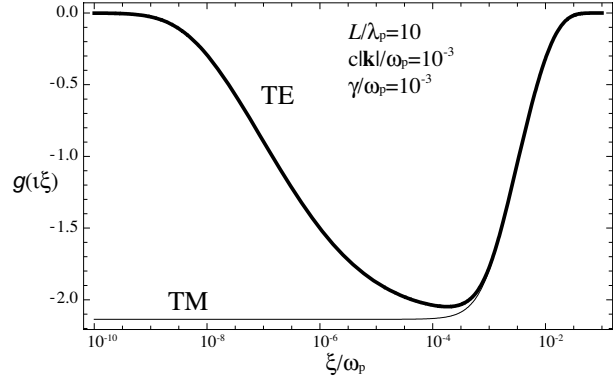}}
  \caption{Behavior of $g(\imath\xi)$ as function of imaginary frequency $\omega = {\rm i}Ê\xi$ for the polarizations TE and TM. The  Drude dielectric function is used. See the text for more detail.}\label{gtetm}
\end{figure}

Let us consider now the limit $\tau\rightarrow 0$, still ignoring 
the explicit temperature dependence of the reflection 
coefficients.
The sum over $k$ in Eq.(\ref{eq:entro}) is approximated by its first
term. We then find that the entropy goes 
to zero:
\begin{equation}
\label{entro}
\tau\rightarrow 0: \quad
S \to -
\left(\frac{1}{2}
\oint_{C}\frac{g(z)}{z}\frac{dz}{2\pi\imath} 
- \frac{1}{2}g(0)\right) = 0
\end{equation}
This result is the same as the one obtained by B\"{o}strom and 
Sernelius \cite{Bostrom00, Bostrom04} when the dissipation rate in
the Drude model does not depend on temperature. 
The present demonstration has the 
advantage to apply to a broad class of reflection 
coefficients that do not depend on temperature, of course. 
The first terms of the expansion in $\tau$ are given by 
	\begin{equation}
	\fl S\approx-\sum_{p,k}\oint_C g(z) 
	\left[-\frac{1}{6}\frac{\imath\tau}{z^2}+\frac{5}{6}\frac{\tau^{2}}{z^{3}} \right]
	\frac{dz}{2\pi\imath}=\sum_{p,k}
	\left[\frac{g_{\xi}(0)}{6} \tau+\frac{5}{12}g_{\xi^{2}}(0)\tau^{2}\right]
	\end{equation}
where we express the derivatives in terms of the imaginary frequency $\xi =-\imath\omega$. In the case of the Drude model, fig.\ref{gtetm} shows that for the TM polarization, 
$g(\imath\xi)$ monotonously increases with $\xi$. Hence, $g_\xi(0)$ 
is close to zero or positive.
The TE polarization behaves differently: it starts from a value 
$g(0)$ close to zero and shows first a negative slope before
increasing and going to zero as $\xi \to \infty$; this implies a negative contribution
to the entropy. This behavior is due to the vanishing of the reflection coefficient 
$r^{\rm TE}_{\mathbf{k}}( \omega \to 0 )$ for any finite $\mathbf{k}$.
In the general case, the sign of the low-temperature
entropy depends of course on the balance between the derivatives
of $g$ for both polarizations; it is therefore model-dependent. 

Nevertheless, we can make some general considerations, which are valid for a large class of metals at least as long as the local description for the reflection coefficients is applicable.
At high frequencies, as long as they are below the plasma frequency, 
both TM- and TE-reflection coefficients are close to unity because 
the dielectric function is large, giving a good 
approximation to the perfect conductor limit $\varepsilon = \infty$.
This behaviour continues down to zero frequencies in TM-polarization
because here, electric fields are dominant and are efficiently 
screened by the (Drude) metal. Only for frequencies much larger than 
plasma frequency $\omega_{\rm p}$, the reflection coefficient goes quickly to zero.
Hence, we expect a small and positive entropy contribution from the  
TM mode at low temperatures. In TE-polarization, the ultraviolet 
behavior is the same, but at low frequencies the field is prevalently magnetic and penetrates 
into the material causing that
the reflection coefficient rapidly decreases to zero \cite{Svetovoy05}.
This strongly suggests that the TE contribution to the Casimir 
entropy is negative and larger in magnitude than its TM counterpart.
Our analysis also explains why we have always a positive entropy
for the plasma (non dissipative) model:
the UV limit is unchanged, 
the behavior of both reflection 
coefficients is qualitatively the same, and both 
do not vanish at $\omega \to 0$. A metal described through the plasma model 
thus behaves qualitatively similar to a superconductor at 
sufficiently small frequencies.
We can conclude that a negative value of the Casimir 
entropy is a general feature of the Lifshitz formula 
when the TE reflection coefficient goes quickly and monotonously to zero in 
the limit of zero frequency.
%------------------figure path-------------------------------------
%----------------------end figure path--------------------------

Let us consider now a situation where the reflection coefficients 
$r( \omega, \tau )$
depend on temperature. In this case it could happen that the
limits $\tau \to 0$ and $\omega \to 0$ do not commute. More 
specifically,
\begin{equation}
\lim_{\tau\rightarrow 
0}r(\alpha\tau,\tau)\not=\lim_{\tau\rightarrow0}r(0,\tau)\Rightarrow 
\lim_{\tau\rightarrow 0}g(\alpha\tau,\tau)\not=\lim_{\tau\rightarrow 
0}g(0,\tau)
\end{equation}
where $\alpha$ is some non-zero complex constant. 
This means that $g( \omega, \tau )$ is no longer an analytic
function in the whole upper half complex plane when $\omega$ and 
$\tau$ are not independent and 
we approach zero along the path $\omega = \alpha\tau$.
To illustrate this with more detail, let us consider 
the Drude model and the Fresnel form of the reflection 
coefficients: 
\begin{equation}
\label{eq:Drude-eps}
\varepsilon(\omega,\tau) = 1 - \frac{\omega_{\rm p}^2}{\omega^2[1 + 
\imath\gamma(\tau)/\omega]}
\end{equation}
where $\gamma( \tau )$ is the
(temperature-dependent) dissipation rate.
The quantity that plays a crucial role is 
$\omega^2\epsilon(\omega,\tau)$ for which it is easy to see that
\begin{equation}
\lim_{\tau\rightarrow 0} (\alpha\tau)^2\epsilon(\alpha \tau,\tau)
\not= \lim_{\tau\rightarrow 0}
\lim_{\omega\rightarrow 0} \omega^2\epsilon(\omega,\tau)
\end{equation}
if
\begin{equation}
\lim_{\tau\rightarrow 0}\gamma(\tau)/\tau=0
\end{equation}
This happens for the example considered in 
Ref.\cite{Bezerra04} where at 
low temperatures, $\gamma(\tau)\propto\tau^2$. 
In this particular case, $\lim_{\tau \to 0} r(\alpha\tau,\tau)
= r_{\rm plasma}(0)$.
As an immediate consequence, we can, in the general case, 
no longer perform the previous manipulations
for all $n$. In fact, we have to treat $n = 0$ separately. 

The problem can be handled defining a new function $\tilde{g}$ which 
is identical to $g$ for all nonzero Matsubara frequencies,
but does not feature the discontinuity discussed above. 
In the case of the Drude model this can be easily done with the
substitution $g(0,\tau) \mapsto
\tilde{g}(0) = \lim_{\tau\rightarrow 0}g(\alpha\tau,\tau)$. The 
function $\tilde{g}$ so defined has the same analytical 
properties as a function  $g$ with no temperature dependence, and 
we can therefore repeat the previous analysis. The substitution has to 
be made without quantitatively changing our expressions. This means 
that we have to rewrite the expression for the entropy given in 
Eq.(\ref{eq:entroform}) as follows 
\begin{equation}
\label{entroform}
\fl S=-
\left[
\frac{ g(0, \tau ) - \tilde{g}(0) }{ 2 }
+\psum\limits_{n=0}^{\infty}\ 
\left[
\tilde{g}(\imath n \tau,\tau)+\imath n\tau \tilde{g}_\omega(\imath 
n\tau, \tau)+\tau\tilde{g}_{\tau}(\imath n \tau,\tau)
\right]
\right]
\end{equation} 
For the first two terms in the sum we can repeat the previous 
analysis and show that they vanish in the limit $\tau\rightarrow 0$. 
The last term in the sum can be approximated by
\begin{equation}
\fl \psum\limits_{n=0}^{\infty}\  \tau\tilde{g}_{\tau}(\imath n 
\tau,\tau)=\partial_{\tau}\gamma(\tau)
\psum\limits_{n=0}^{\infty}\ 
\tau \frac{ \partial\tilde{g}(\imath n 
\tau,\tau) }{ \partial \gamma}
\stackrel{\tau\rightarrow 0}{\longrightarrow}
\partial_{\tau}\gamma(\tau)
\int\limits_0^{\infty}
\frac{ \partial\tilde{g}(\imath\xi,\tau) }{ \partial \gamma}
d\xi
\end{equation}
which is equal to zero for an at least quadratic temperature dependence
of $\gamma(\tau)$. In this case the low-temperature limit of the
Casimir entropy is 
\begin{equation}
    S = - \sum_{p,\mathbf{k}}
\frac{ g(0) - \tilde{g}(0)}{ 2 } = 
\frac12 \sum_{\mathbf{k}}\tilde g^{\rm TE}_{\mathbf{k}}(0) < 0
\end{equation}
where we have restored the sum over the wavevectors.  
Inspection shows that
this is nothing but the result of Bezerra et al.\cite{Bezerra04}.

%%%%%%%%%%%%%%%%%%%%%%%%%%%%%%%%%%%%%%%%%%%%%%

\section{A sum over modes in the dissipative case}

One of the difficulties in generalizing Eq.(\ref{eq:CasEn}) to a
dissipative system is the concept of a ``mode'' because this
represents a state of constant energy which can be well defined only
for a closed system where dissipation does not occur.  The naively
constructed complex mode frequencies actually do not represent the
real modes of our system.  In fact, in quantum mechanics a dissipative
system is by definition an open
one~\cite{Caldeira81,Caldeira83,Breuer02,Ford65,%
Ford87a,Intravaia03}, i.e.\ a system coupled a macroscopic system,
generally a thermal bath.  The bath contains a continuum of degrees of
freedom into which the system energy can be irreversibly dissipated.
A formula like Eq.(\ref{eq:CasEn}) is applicable only if we consider
the modes of the total system---which is, of course, cumbersome or
plainly impossible in practice \cite{Sernelius06}.  It is natural to
ask, however, whether it may be useful to apply the familiar picture
of a damped oscillator in terms of a complex frequency, which is just
the pole of the Green function for this system.  Here we are going to
show that this compact description can be applied to the Casimir
energy as well, with the complex poles going over into conventional
modes in the non-dissipative case.

\subsection{Sum over poles}

Our starting point will be the Lifshitz formula for 
the Casimir force between two slabs with finite thickness 
\cite{Lifshitz56,Jaekel91,Genet02,Genet03}
\begin{equation}
\label{f1} 
F = \frac{\hbar }{4\pi} \sum_{p,\mathbf{k}} 
\partial_L 
\int_{-\infty}^{\infty}d\omega\ 
\coth\left[\frac{\pi\omega}{\tau}\right]
\im{\ln\nt{D}( \omega )}^L_{\infty}
\end{equation}
where the frequency integral is just above the real axis (${\rm Im}\, 
\omega > 0$) ensuring retarded response functions.
The environment is at the scaled temperature $\tau$ (see 
Eq.(\ref{eq:scaled-tau})). We recall the renormalization,
$[\cdots]^L_{\infty}$, with respect to 
a pair of infinitely distant mirrors.
In the following, we drop again the sum over $p$ and
$\mathbf{k}$.
Finally, the relevant dispersion function is
\begin{equation}
\partial_L 
\ln{D}( \omega ) 
= \kappa\frac{1 + r(\omega)^2\, {\rm e}^{-2 \kappa L}}
             {1 - r(\omega)^2\, {\rm e}^{-2 \kappa L}},\quad
\end{equation}
with $r(\omega)$ being the reflection coefficient of 
the mirrors (supposed identical for simplicity).  
We used the reality of the optical response,
${r}(\omega)^* = {r}(-\omega^*)$, to extend the integral~(\ref{f1}) 
to the negative frequency axis. 
After a few manipulations, the Casimir \emph{energy} can be written
\begin{equation}
\label{energy}
E=\frac{\hbar}{2}\sum_{p, \mathbf{k}} \nt{\Omega},\quad %XX]
{\Omega}=-\int_{-\infty}^{\infty} \frac{d\omega}{2\pi}\ \omega 
\coth\left[\frac{\pi\omega}{\tau}\right] \im{
\partial_{\omega} \ln {D}( \omega ) }^{L}_{\infty}
\end {equation}
The integrand of this equation goes to zero sufficiently fast
for $\abs{\omega}\rightarrow\infty$ so that we can close the integral in the complex plane along a path which encloses the lower half-plane in the clockwise sense.. 

The function $D$ is constructed in such a way that its
zeros are the complex ``mode'' frequencies we 
are looking for. But, $D$ also shows branch cuts along the real axis
that prevent a straightforward deformation of the integration contour.
These branch cuts come in two types: the first 
corresponds to the mode continuum incident from outside 
the cavity; the second
is connected with the dissipative phenomenon and arises
from the continuum of modes describing the bath. 
Following Ref.\cite{Schram73}, we eliminate the branch 
cut of the first type by placing our system into a 
perfectly reflection box with a length $a>L$. The second type
of cut is avoided by 
analytically continuing $D$ from the upper half to the lower half
of the complex frequency plane.  It is exactly because of this 
step that the zeros of $D$, otherwise located on a 
unphysical sheet, become apparent \cite{Sernelius06}.
With these changes, $D$ 
turns into a function $D_{a}$ that shows, 
instead of a branch cut, 
a set of discrete zeros  that adds to 
the ones of ${D}$. One can proceed in the usual way with $D_{a}$ and
take the limit $a\rightarrow\infty$ at the end of the calculation.

All functions involved in the integral~(\ref{energy}) are now
meromorphic functions \cite{Markusevic88} and 
can be expanded over their singularities as follows
\begin{equation}
\fl \coth\left[ \frac{\pi\omega}{\tau}\right]= 
\sum_{n=-\infty}^{\infty}
\frac{\tau / \pi}{\omega + \imath n\tau},
\quad
\partial_{\omega} \ln D_{a}( \omega ) = \sum_{m}\left(
\frac{1}{\omega-\omega_{m}}+\frac{1}{\omega+\omega_{m}^{*}}\right)
\label{poleexpansion}
\end{equation} 
The zeros of $D_{a}$ must satisfy ${\rm Im}\, \omega_{m} < 0$
because of causality and passivity. Moreover, the dispersion function
inherits
the reality condition, $D_{a}( \omega )^* = D_{a}( -\omega^* )$,
of the 
reflection coefficients. Therefore, the zeros occur in pairs,
$\omega_{m}$ and $-\omega_{m}^*$, on both sides of the imaginary axis. 
 
With a little algebra and exploiting the residues theorem one can 
show
\begin{equation}
\label{eq:ene}
\fl \Omega=\re{\sum_{m}\omega_{m}
\coth\left[\frac{\pi\omega_{m} 
}{\tau}\right]}^{L}_{L\rightarrow\infty}
+\re{\sum_{m} \sum_{n=0}^{\infty}
\frac{ \imath \omega_{m} }{ \pi }
\frac{2n\omega^{2}_{T} }{ (n\tau)^{2}+\omega^{2}_{m}}
}^{L}_{\infty}
\end{equation}
where we have already taken the limit $a\rightarrow\infty$. 
Note also the following identity
\begin{equation}
\label{depe}
\im{\sum_{m}\omega_{m} }^{L}_{\infty}=
\int_{-\infty}^{\infty}d\omega\ \omega\  
\im{ \partial_{\omega} \ln{D} }^{L}_{\infty} = 0
\end{equation}
than can be easily shown because of parity.
This sum rule ensures that in the sum over $n$ 
in Eq.(\ref{eq:ene}), the summands 
go like $1/n^{2}$ for $n\rightarrow \infty$.  

As a check let us consider the non-dissipative case: 
the frequencies $\omega_{m}$ are then real, the second term
in Eq.(\ref{eq:ene}) 
vanishes, and the Casimir energy can be written as
\begin{equation}
    E = \frac{ \hbar }{ 2 } \sum_{p, \mathbf{k}}\left[ 
\sum_{m}\omega_{m}\left(
2 \bar{n}_{T}(\omega_{m}) + 1\right)
\right]^{L}_{\infty}
\end{equation}
where $\bar{n}_{T}(\omega_{m})$ is just the mean 
number of thermal photons in the frequency mode $\omega_{m}$. 
In the 
limit of zero temperature this number goes to zero and we get the 
expressions of Ref.\cite{Intravaia05, Intravaia07}. 

In the dissipative case, we focus on 
the zero temperature limit ($\tau\rightarrow 0$): the 
hyperbolic cotangent in Eq.(\ref{eq:ene})
goes to unity, while the second term becomes an integral so that 
we have
\begin{equation}
\label{finaldissene}
\tau \to 0: \quad
{\Omega} \to 
{\rm Re}\,\bigg[\sum_{m}\omega_{m} \bigg]^{L}_{\infty} 
+ 
\int\limits_{0}^{\infty}d\xi\ 
{\rm Re}\,\bigg[ \sum_{m}\frac{\imath\omega_{m}}{\pi}
\frac{2\xi}{\xi^{2}+\omega^{2}_{m}}
\bigg]^{L}_{\infty}
\end{equation}
The integral converges at the upper limit thanks to the sum 
rule~(\ref{depe}) and we can write
\begin{equation}
\label{finalresult2}
E = \frac{\hbar}{2} \sum_{p,\mathbf{k}} {\rm Re}\,\bigg[\sum_{m}
\big(\omega_{m}-
\frac{2\imath\omega_{m}}{\pi}
\ln\frac{\omega_{m}}{\Lambda}
\big)\bigg]^{L}_{\infty}
\end{equation}
where the meaning of the constant frequency $\Lambda$ will become
clear soon.
This is the main result of this section: we have found the Casimir 
energy between dissipative mirrors, at zero temperature,
as a sum over cavity mode frequencies.
The expression is valid for a generic (causal)
dielectric function and for mirrors with arbitrary thickness.
One sees immediately that the non-dissipative case reduces to the 
standard sum over real mode frequencies.

To interpret the two terms in Eq.(\ref{finalresult2}),
observe that for
each collection of quantum numbers $p,\mathbf{k},m$, we deal with 
the average energy of a quantum
damped harmonic oscillator, as discussed in detail by Nagaev and 
B\"{u}ttiker~\cite{Nagaev02}. The first term of the energy 
is the familiar zero point energy, renormalized by 
the coupling to the environment (see also Ref.\cite{Intravaia03}). 
The second part of the energy is the most interesting one. It arises
because
the ground state of the oscillator is no longer an eigenstate of the 
whole oscillator+bath system. (This can be traced back to 
antiresonant terms in the coupling Hamiltonian.) As a consequence,
the oscillator energy is subject to quantum fluctuations, whose 
average leads to an extra energy with a logarithmic behavior.
Within this analogy, the constant $\Lambda$ takes the role of a cutoff 
frequency for the coupling to the bath. In fact, for a single 
oscillator, the cutoff is needed to renormalize some quantities like 
the momentum fluctuations~\cite{Nagaev02}.
It is interesting to note that for the Casimir energy, 
the sum rule~(\ref{depe}) removes the cutoff dependence 
in the final result.

\section{Conclusions}

We have explored the impact of dissipation and of temperature on the 
Casimir effect. On fairly general grounds, we have 
demonstrated that the Casimir entropy vanishes at $T \to 0$
(the Nernst theorem is satisfied) when the material 
properties (reflection coefficients) behave continuously as both 
frequency and temperature are decreased to zero. A discontinuity in 
the plane of these variables leads to a non-vanishing, typically negative entropy. We 
have argued that this can be traced to the behavior of the TE 
reflection coefficient at low frequencies. 
Both results are in agreement with previous work 
addressing specific models. The complex mode frequencies that occur 
between dissipative slabs are finally a useful tool to compute the 
Casimir energy as a generalized sum over ``modes'', although the mode 
concept has to be treated with care in this intrinsically open system. 
Our expression Eq.(\ref{finalresult2}) shows that the Lifshitz formula (that we started with) does not reduce to the ``recipe'' of taking the real part of a complex sum over modes, as stated in Refs.~\cite{Langbein70, Langbein74}. The latter sum deeply changes its structure if dissipation is allowed for, while the Lifshitz expression remains formally the same. We have argued that the logarithmic correction that we find has a clear meaning from a physical point of view, by drawing the analogy to a damped quantum oscillator. As it happens, the result is 
independent of the cutoff frequency for the coupling to the environment.

\section{Acknowledgments}

F. Intravaia thanks the Alexander von Humboldt Foundation for 
financial support.

\end{document}